\documentclass[12pt,a4paper,onecolumn,doublespacing,usenatbib]{mnras}
\usepackage{amsmath}
\usepackage{natbib}
\title{Realistic coasting  cosmology from the Milne model}
\author[Moncy V. John]{Moncy V. John  \thanks{moncyjohn@yahoo.co.uk}\\
Department of Physics, St. Thomas College, Kozhencherri - 689641, Kerala, India}

\pubyear{2016}

\date{\today}

\begin{document}
\label{firstpage}
\pagerange{\pageref{firstpage}--\pageref{lastpage}}
\maketitle

\begin{abstract}
In the context of the recent synchronicity problem in $\Lambda$CDM cosmology, coasting models such as the classic Milne model and the $R_h=ct$ model have attracted much attention. Also, a very recent analysis of supernovae Ia data is  reported to favour   models with constant expansion rates. We point out that the nonempty $R_h=ct$ model has some known antecedents in the literature. Some of these are published even before the discovery of the accelerated expansion and were shown to have none of the cosmological problems and also that $H_0t_0=1$ and  $\Omega_m/\Omega_{dark \; energy}$ = some constant of the order of unity. In this paper, we also derive such a model by a complex extension of scale factor in the Milne model.

\end{abstract}


\section{Introduction}

The synchronicity problem  \citep{avelino2016} is the most recent among a series of problems that cropped up  in the $\Lambda$CDM  cosmology during the past few decades. Several   problems were identified in cosmology during the 1980s, but the theory of inflation \citep{guth1981} solved most of them at one stroke. This theory, which speculates quantum field theoretical effects in the early epochs,  suggests that the present  universe is flat. But  when the condition of flatness is combined with the measured high value of the Hubble parameter,  there arose  an `age problem'  in the middle of the 1990s. The discovery of the accelerated expansion in 1998 predicted  a cosmological constant $\Lambda$ or some unknown dark energy   in the present universe. This discovery resolved  the age problem, but in its turn,  seeks an explanation to the near equality of the energy density corresponding to such $\Lambda$ or dark energy and the matter density in the present universe. This is the `coincidence problem' in cosmology, for which the $\Lambda$CDM model offers no  satisfactory solution  yet. Recently, three important cosmological observations, namely, the apparent magnitude and redshift  of  Type Ia supernovae (SN Ia), CMB power spectrum, and baryon acoustic oscillations, together predict that the dimensionless age $H_0t(a)$ of the present universe is very close to unity \citep{avelino2016}. This  is  puzzling since according to the $\Lambda$CDM model,   the product  $H_0t$ could have values very different from unity; it can be anywhere in the range  $0<H_0t(a) < \infty$. Except during the period of inflation,  in the past or future of the universe, this value shall not be unity either.  Just like the coincidence problem,  this synchronicity problem in the $\Lambda$CDM model  is also regarding the  special status ascribed to the epoch in which we live today. 

 On the other hand, the classic Milne model \citep{milne1935}   has no synchronicity problem. When  viewed as a Friedmann model,  its  scale factor  obeys $a\propto t$.   In this case, the present dimensionless age of the universe $H_0^{-1}$ would exactly correspond to the actual age $t_0$, and this result will  always remain valid.
Hence in this cosmological model,  $H_0t_0=1$ is not a problem; instead, this is its prediction.  In \citep{avelino2016}, it is mentioned that the Milne model is the most suitable one to explain the above observational result, but it is soon rejected  on grounds that the   model is empty and has negatively curved space sections (i.e.,  $\rho=0$ and $k=-1$). This is quite reasonable, for an empty model is not a realistic one. 

Very recently, an analysis of Type Ia supernova data  \citep{nielsen2015marginal}  has  appeared with the result that there is only  marginal  evidence for the widely accepted claim of the accelerated expansion of the universe.  By a rigorous statistical analysis using the Joint Lightcurve Analysis (JLA) catalogue  of 740 SN Ia, it is found that the SN Ia Hubble diagram appears consistent with a uniform rate of expansion. This brings  the Milne model  again to the centre-stage of cosmology, albeit in some new avatar. 

Recently, an $R_h=ct$ cosmological model \citep{melia2012rh}, which has an expansion history coinciding with that of the Milne model, is suggested as the true model of the universe \citep{melia2012fitting,melia2013cosmic}. It is claimed that one-on-one comparative tests carried out between this model and the $\Lambda$CDM model using over 14 cosmological measurements and observations give  conclusive evidence in support of the $R_h=ct$ model \citep{melia2015recent}. In the context of the latest synchronicity problem in cosmology and the analysis of JLA catalogue mentioned above, this model assumes great significance. Possibly due to  this, in the literature, there are growing concerns  regarding the fundamental basis of the theory itself \citep{bilicki2012we,mitra2014rh}. For instance, the latest paper on this subject \citep{mitra2014rh}  contends that $R_h = ct$ is a vacuum solution. Addressing this criticism, it is shown in \citep{melia2015recent} that the $R_h=ct$ model is  not empty and hence not the same as the Milne model.

In view of the mounting supportive evidences obtained from the above analyses of most recent cosmological data \citep{avelino2016,nielsen2015marginal,melia2015recent}, one notes that models such as the above deserve continued scrutiny, both from the conceptual and observational fronts. In this paper, we first point out that  there exist some  antecedents \citep{john1996modified,john1997low,john2000generalized} to the $R_h=ct$ cosmological model, though they  were not referenced in \citep{nielsen2015marginal,melia2012rh,melia2012fitting,melia2013cosmic,melia2015recent}. As in the $R_h=ct$ model, the expansion   in these nonempty, real models is `always coasting' (i.e., in accordance with scale factor $a\propto t$, except possibly during or immediately after the Planck epoch) and this is  due to the vanishing of gravitational charge $\rho+3p$, rather than the vanishing of $\rho$. The most closely related antecedent is the one  proposed in \citep{john2000generalized} (where $a=ct$ for $k=0$). However, its difference in evolution history with that in \citep{john1996modified,john1997low} is only around  the Planck epoch and is insignificant at the observational front. The vanishing of cosmological problems \citep{john1996modified,john1997low,john2000generalized} and comparison of this model with the $\Lambda$CDM model using the then available SN Ia data \citep{john2002comparison,john2005cosmography,john2010bayesian} were discussed at length.   It is unfortunate that such known  published works are ignored in the recent literature on the possibility of constant expansion rate of the universe. 

Following this discussion,    in this paper we also show that  the Milne model can  lead to  a  nonempty, coasting cosmological model. This is done by a complex extension of scale factor in the former model. If  we  take the energy density in the nonempty model  as the sum of matter density and a time-variable dark energy density and  do not assume any arbitrary conservation law for these individual components,  the coincidence problem can be seen to vanish in it. The new  model is realistic and nonsingular and is always coasting after the Planck epoch.   As a next step, a quantum cosmological treatment of this model is made, which provides the result that the minimum radius $a_0$ in it is of the order of the Planck length. The complexified Milne model has  Eucledean (++++) signature at $t=0$, but  it changes to the usual Lorentzian one (+\;-\;-\;-)  immediately after the Planck epoch. We find that this leads to the widely speculated `signature change' \citep{ellis1992change,hayward1992signature} in the early universe. 

The organisation of this paper is as follows. In Sec. 2, we discuss the antecedents to the $R_h=ct$ model and in Sec. 3,  a new derivation of the flat, nonempty coasting model is presented. The quantum cosmological treatment of this model is made in Sec. 4.    The last section comprises a discussion of the results.

\section{Antecedents to the $R_h=ct$ model}

 In a well-known work,  \cite{kolb1989coasting} has discussed at length the implications of a coasting cosmological model. In this, the universe  is presently dominated by some exotic K-matter and is coasting, with equation of state $p_K=-\rho_K/3$. However, the model is different from the $R_h=ct$ or other  `always coasting' models, for during  the major part of its history in early epochs, the universe is dominated by radiation and matter and hence its expansion rate is not uniform. 
 
 A nonempty, closed cosmological model that is  coasting throughout the history of the universe  after the Planck epoch  was first  proposed in 1996 \citep{john1996modified,john1997low}. This model coincided with  the widely discussed Ozer-Taha model \citep{ozer1986possible,ozer1987model} at the earliest epochs.   The attempt by Ozer and Taha was to put forward an alternative to the theory of inflation, which solves the flatness problem, horizon problem, etc. They obtained   a bouncing,  nonsingular  solution with $a=\sqrt{(a_0^2+t^2)}$, and speculated that $a_0$, the minimum radius of the universe, has some small value. After the bounce, the  model reaches the $a\propto t$ phase for some time, but soon deviates from it to enter the decelerating standard big bang evolution and continued in it.  On the contrary, the  antecedent to the $R_h=ct$ model mentioned above in \citep{john1996modified,john1997low} has the  evolution $a=\sqrt{(a_0^2+t^2)}$ throughout the cosmological history. Moreover, a quantum cosmological treatment gave the result that the minimum radius $a_0$ is of the order of Planck length.  These models  were shown to have none of the cosmological problems, such as singularity, horizon, flatness, monopole, cosmological constant, size, age, etc. 
 
The more closely related antecedent in \citep{john2000generalized} is an always coasting cosmology with $a\propto t$, obtained  by extending the dimensional argument of  \cite{chen1990implications}. This modified Chen-Wu model, which can have $k=0,\pm 1$,   has both matter and  a time-varying cosmological constant. Comparison of this model with the $\Lambda$CDM model was performed using the SNe Ia data,  with the help of  the Bayesian theory \citep{john2002comparison}. The results showed that the evidence against the coasting model, when compared to the $\Lambda$CDM model, is only marginal. In 2005, an analysis was again performed  \citep{john2005cosmography} using the then available SNe data to see whether the data really favours a decelerating past for the universe. Again, the  conclusion was that the evidence  is not strong enough to discriminate this case from a coasting cosmology. Recently, a quantum cosmological treatment \citep{john2015exact} showed that  a coasting solution is unique, since it has identical classical and quantum evolution, as in the case of free particles described by plane waves in ordinary quantum mechanics. The $R_h=ct$ model coincides  with the coasting model in \citep{john2000generalized}, for $k=0$.

Some other realistic coasting models have also appeared  in the literature since then.  Nucleosynthesis in a different coasting model was discussed by Lohiya and co-workers in \citep{sethi1999comment,lohiya1999programme,batra2000nucleosynthesis}. This issue is  pursued  in several works, including some recent ones \citep{singh2015constraints}. They have also investigated the status of realistic coasting models with regard to other cosmological observations, such as the SNe Ia data \citep{dev2001linear}. 

 The synchronicity problem does not arise in these coasting models  due to their unique expansion history $a\propto t$.  Here it is   noted that such a model has vanishing gravitational charge $\rho +3p$, or equivalently an equation of state $p=-\rho/3$. As described explicitly in the  antecedents \citep{john1996modified,john1997low,john2000generalized}, the model is devoid of any coincidence problem too, when a proper accounting of its energy densities is made. One notes that the Einstein equation in general relativity implies a conservation law for energy-momentum, but  this is valid only for the total energy density. If we assume that the cosmic fluid comprises of matter and a vacuum energy, and if we do not assume any arbitrary  conservation law for the individual components, then there can be some interaction between them. In the present case, this may lead to creation of matter from vacuum energy, as can be seen from the fact that here both energy densities vary as $a^{-2}$. (However, the creation rate shall be so small that it would be impossible to detect such events at the present level of precision in observations.) Thus the coasting model, when assumed to contain a decaying vacuum energy,  naturally leads to a constant ratio between the energy densities of matter and vacuum, and this ratio shall be of the order of unity. That this resolves the coincidence problem  is noted also in \citep{melia2016epoch}. 

\section{Complex extension of the Milne model} \label{sec:milne}

Let us now  extend the scale factor  of the Milne model to the complex plane  and denote it  as $ \hat{a}$. 
 We can now show that this results in  a coasting  cosmology with real scale factor $a\equiv \mid \hat{a}\mid$. One can write the Friedmann equations corresponding to the empty Milne model   as

\begin{equation}
\left( \frac { \dot {\hat {a}}}{\hat {a}} \right) ^{2} - 
\frac {1}{\hat {a}^{2}} = 0 \label{eq:t-tc}
\end{equation}
and

\begin{equation}
2\frac {\ddot {\hat {a}}}{\hat {a}} + \left( 
\frac {\dot {\hat {a}}}{\hat {a}}\right) ^{2} - \frac {1}{\hat
{a}^{2}} = 0. \label{eq:s-sc} 
\end{equation}
 Using the polar form $\hat{a}=ae^{i\phi}$ in these equations and then equating their real  parts, we get

\begin{equation}
\frac {\dot {a}^{2}}{a^{2}}  = \dot {\phi }^{2} +
\frac {1}{a^{2}} \cos 2\phi , \label{eq:t-tphi}
\end{equation}
and

\begin{equation}
2\frac {\ddot {a}}{a}+\frac {\dot {a}^{2}}{a^{2}} 
= 3 \dot {\phi }^{2} + 
\frac {1}{a^{2}} \cos 2\phi  . \label{eq:s-sphi}
\end{equation}
These appear as the Friedmann equations for a flat ($k=0$) model, filled with a homogeneous scalar field $\phi$. This model is quite different from that of Milne, since the geometry of its space sections and the energy density are different. The right hand sides of the equations imply the presence of a scalar field with kinetic energy $\dot {\phi }^{2}$ and potential energy $\cos(2\phi)/a^2$. That is, we have a new cosmological model with real scale factor $a=\mid \hat{a} \mid$, whose space sections are flat  and which is nonempty. Equating also the imaginary parts, we get two supplementary equations

\begin{equation}
\ddot {\phi } + 2\dot {\phi }\frac {\dot {a }}{a} =0,
\label{eq:consphi} 
\end{equation}
and

\begin{equation}
2\dot {\phi }\frac {\dot {a}}{a} = -\frac {1}{a^{2}}\sin 2\phi
.\label{eq:phidot} 
\end{equation}
These shall be of help in solving the  system of equations (\ref{eq:t-tphi}) and (\ref{eq:s-sphi}). However, one can directly  solve equations  (\ref{eq:t-tc}) and (\ref{eq:s-sc}) to obtain $\hat{a}=(c_1 \pm ct)+ic_2$. Choosing the origin of time such that the constant of integration $c_1=0$ and relabelling $c_2\equiv a_0$, we get the solution for $\hat{a}$ as

\begin{equation}
\hat{a}= \pm t+ia_0. \label{eq:comp_soln}
\end{equation}
 The solution for $a=\mid \hat{a} \mid$  can be seen to be 

\begin{equation}
a=\sqrt{a_0^2+t^2}, \label{eq:real_a}
\end{equation}
and the argument $\phi$ can be obtained as

\begin{equation}
\phi =\tan^{-1}\left(\frac{a_0}{t}\right). \label{eq:real_phi}
\end{equation}
This is a bouncing, nonsingular evolution as in the early phase of Ozer-Taha model \citep{ozer1986possible,ozer1987model}, with $a_0$ as the minimum value for the scale factor. The time at which this minimum occurs for $a$ is taken as $t=0$.

The energy density $\rho$ and pressure $p$ in the new model, as can be deduced from equations (\ref{eq:t-tphi}) and (\ref{eq:s-sphi}), have the following evolution.  At the epoch near $t=0$, the kinetic energy of the field $\phi $ is dominant, leading to the  nonsingular behaviour. This phase of evolution is capable of solving cosmological problems such as that related to the horizon. For large $t$, we have $\rho \propto a^{-2}$ and $\rho +3p \approx 0$. The contribution for $\rho$ in the late universe comes almost entirely from the potential energy term. It may also be noted that its magnitude is very nearly equal to the critical density for the universe. One can explicitly write the variation of  total density and pressure with  scale factor  as

\begin{equation}
\rho=\frac{3}{8\pi G}\left(\frac{1}{a^2}-\frac{a_0^2}{a^4}\right)
\end{equation}
and

\begin{equation}
p=-\frac{1}{8\pi G}\left(\frac{1}{a^2}+\frac{a_0^2}{a^4}\right),
\end{equation}
so that

\begin{equation}
\rho + 3p=-\frac{3}{4\pi G} \frac{a_0^2}{a^4}.
\end{equation}
The term that causes  the bounce to happen corresponds to a negative energy density, which can now be separated as 

\begin{equation}
\rho_-=-\frac{3}{8\pi G}\frac{a_0^2}{a^4} \label{eq:neg_energy}
\end{equation}
The pressure due to this is given by $p_-=(1/3)\rho_-$, which is an equation of state characteristic of relativistic energy densities. This energy density becomes negligible for $a\gg a_0$, when compared to the rest of the energy densities.

  Thus  the  real  model  is flat, with the total energy density and pressure obeying $\rho+3p\approx 0$  for $a\gg a_0$. After this initial epoch, if the potential energy of the field  comprises energy corresponding to  radiation/matter and a time-variable dark energy,  we can write $\rho =\rho_m+\rho_{d.e.}$. Taking $p_m=w\rho_m$ and $p_{d.e.}=-\rho_{d.e.}$, one obtains 
  
\begin{equation}
\Omega_m\equiv \frac{\rho_m}{\rho_c}=\frac{2}{3(1+w)}, \qquad \Omega_{d.e.}\equiv \frac{\rho_{d.e.}}{\rho_c}=\frac{1+3w}{3(1+w)}.
\end{equation}  
Since this universe is flat, the total density parameter $\Omega=\rho/\rho_c =1$. When matter in the present universe is   considered to be nonrelativistic with $w=0$, the above equations     predict $\Omega_m =2/3$ and $\Omega_{d.e.}=1/3$. The ratio between these densities is a constant of the order of unity throughout the expansion history (for $a\gg a_0$),  and this avoids the coincidence problem. Being a coasting evolution, it naturally has no synchronicity problem. It  solves all other cosmological problems, as demonstrated in the previous works.

With the aid of the solution  (\ref{eq:comp_soln}), one can see that the complex-extended Milne model has  Eucledean (++++) signature at $t=0$, but  it changes to the usual Lorentzian (+\;-\;-\;-) one for $a\gg a_0$. Hence the complex extension of Milne model leads naturally to a signature change in the early universe, a possibility discussed extensively in the literature \citep{ellis1992change,hayward1992signature}. The famous Hartle-Hawking `no boundary' boundary condition \citep{hawking1984quantum} in quantum cosmology envisages a change of signature in the early universe. In this regime, there is no time and the spacetime is purely spatial. Even in the classical Einstein equations, it is argued that  the metric is Lorentzian  not because it is demanded by the field equations; instead, it is a condition  imposed on the metric before one looks for solutions \citep{ellis1992change}.  In our case, the complex Milne model undergoes a signature change  in a very natural way. 

\section{Wheeler-DeWitt equation}

In this section, we write down the Wheeler-DeWitt equation for the complex-extended Milne model. For this, we first note that the classical Milne model follows from the Lagrangian

\begin{equation}
L=-\frac{3\pi}{4G}\left(\frac{\dot{\hat{a}}^2\hat{a}}{N}+N\hat{a}\right)
\end{equation}
Here $N$ is called the lapse function, for which one can fix some convenient gauge. Writing down the Euler-Lagrange equations with respect to the variables $N$ and $\hat{a}$ and fixing the gauge $N=1$ leads to equations (\ref{eq:t-tc}) and (\ref{eq:s-sc}), respectively. The canonically conjugate momentum is 

\begin{equation}
\hat{\pi}_{a}=\frac{\partial L}{\partial \dot{\hat{a}}}=-\frac{3\pi }{2G}\frac{\dot{\hat{a}}\hat{a}}{N} \label{eq:can_conj_mom}
\end{equation} 

The canonical Hamiltonian can now be constructed as

\begin{equation}
{\cal H}_c=\hat{\pi}_{a}\hat{a}-L =N\left(-\frac{G}{3\pi }\frac{\hat{\pi}_{a}^2}{\hat{a}} +\frac{3\pi}{4G}\hat{a}\right) \equiv N{\cal H}
\end{equation}

Quantisation of  a classical system like the one above
means introduction of a wave function $\Psi (\hat{a} )$
and requiring that it satisfies \citep{kolb1990early}

\begin{equation}
i\hbar \frac {\partial \Psi }{\partial t} = {\cal H}_{c}\Psi = N{\cal H}\Psi. \label{eq:gwd}
\end{equation}
To ensure that time reparametrisation invariance is not lost at the
quantum level, the conventional practice is to ask that the wave
function is annihilated by the operator version of ${\cal H}$; i.e.,

\begin{equation}
{\cal H}\Psi =0. \label{eq:wd}
\end{equation} 
Eq. 
(\ref{eq:wd}) is called the Wheeler-DeWitt  equation. It is analogous to a zero energy Schrodinger equation, in
which the dynamical variable $\hat{a} $  and its
conjugate momentum $\hat{\pi}_{a} $  
 are replaced by the corresponding operators. The wave
function $\Psi $ is defined on the minisuperspace with just one coordinate $\hat{a}$ and we expect it to
provide information regarding the evolution of the universe. We may note here  that the wave function is independent of time; it is a stationary solution in the minisuperspace. 

The Wheeler-DeWitt equation for our case can be written by making the operator replacements for $\hat{\pi}_{a}$ and $\hat{a}$ in ${\cal H}$. However, finding the operator corresponding to $\hat{\pi}_{a}^2/\hat{a}$ is problematic due to an operator ordering ambiguity. We shall adopt the most commonly used form 

\begin{equation}
\frac{\hat{\pi}_{a}^2}{\hat{a}} \rightarrow -\hat{a}^{-r-1}\frac{\partial}{\partial \hat{a}}\left( \hat{a}^r \frac{\partial}{\partial \hat{a}} \right),
\label{eq:op_order}
\end{equation} 
where the choice of $r$ is arbitrary and is usually  made according to convenience.  Using this expression with $r=-1$, we obtain the Wheeler-DeWitt equation for the complexified Milne model as

\begin{equation}
\frac{d^2\Psi}{d\hat{a}^2} -\frac{1}{\hat{a}}\frac{d\Psi}{d\hat{a}} +  \frac{9\pi^2}{4G^2}\hat{a}^2 \Psi =0.
\end{equation}
This equation has an exact solution

\begin{equation}
\Psi (a)\propto \exp \left( \pm i \frac{3\pi}{4G}\hat{a}^2\right).\label{eq:Psi_coasting}
\end{equation}
Whether this solution corresponds to the classical evolution of the model can be determined by drawing the de Broglie-Bohm trajectories for this wave function \citep{john2015exact}. Identifying $\Psi \equiv R\exp(iS)$, one can draw these quantum trajectories by using the de Broglie equation of motion 

\begin{equation}
\hat{\pi}_a=\frac{\partial S}{\partial \hat{a}}
\end{equation}
In the present case of quantum cosmology, while using (\ref{eq:can_conj_mom}), this equation of motion simply reads

\begin{equation}
-\frac{3\pi }{2G}\dot{\hat{a}}\hat{a}=\pm \frac{3\pi}{2G}\hat{a} \qquad \hbox{or} \qquad \dot{\hat{a}}=\pm 1,
\end{equation}
which is the same classical equation (\ref{eq:t-tc}). Hence, here we have the same classical and quantum trajectories, implying identical behaviour as in the case of free particles described by plane waves in ordinary quantum mechanics.

A notable feature here is the appearance of the factor $\sqrt{{2G}/{3\pi}}$ in the wave function (\ref{eq:Psi_coasting}). This has value very nearly equal to the Planck length. In all quantum gravity theories, a natural length scale is the Planck length. Hence, one can deduce that the value of $a_0$, the imaginary constant appearing in the scale factor of the complex-extended Milne model is also of this value. In turn, this is the minimum radius of the bouncing, real, coasting model. Thus we see that the nonsingular behaviour of the present  coasting model is due to the quantum effects at the earliest moments within the Planck time.

\section{ Discussion} \label{sec:discussion}

The first coasting model for the universe, which was conceived  by Milne, has  been an attractive idea, primarily for its simplicity. Though it is an empty model and is devoid of any gravitational effects, the model is not forgotten even after three quarters of a century.  The nonempty `always coasting' cosmology in \citep{john1996modified,john1997low} was proposed at a time when the universe was considered to be decelerating at the present epoch; i.e., well before the discovery of the unnatural dimming of SN Ia at large redshifts that led to the claim that the universe is accelerating. The solution of the cosmological problems, including the age and coincidence problems, was one of the strong motivations for its prediction. Presently, the model is more relevant in the context of the synchronicity problem, for  the currently popular $\Lambda$-CDM model has no easy way out of it. The former model shall always remain a potential rival to the latter, unless more sophisticated data  shows that $H_0t_0\neq 1$. (As mentioned by \cite{avelino2016}, if this value is  very near to unity and is not exactly equal to it, there would be an even worst sychronicity problem.)

One can see that the complex extension of Milne model, which led to the nonempty, flat model with $a=\sqrt{a_0^2+t^2}$, naturally brings in a negative energy which causes the bounce in the real model, and this is what relieves the model from the singularity problem. As can be expected, this negative energy  has an equation of state corresponding to relativistic matter. There are speculations on a universe driven by a Casimir energy, which is negative  \citep{jaffe2005casimir}.   Casimir first showed, on the basis of relativistic quantum field theory, that between two parallel perfect plane conductors separated by a distance $l$, there is a renormalised energy $E=\pi^2/720l^4$ per unit area \citep{casimir1948attraction}. For a static universe and for a massless scalar field, it was calculated that Casimir energy has a density \citep{elizalde1994vacuum}

\begin{equation}
\rho_{casimir} = -\frac{0.411505}{4\pi a^4}.
\end{equation} 
If we accept the value of $a_0=\sqrt{2G/3\pi}$, which is nearly equal to the Planck length,  our expression for negative energy (\ref{eq:neg_energy}) can be written as

 \begin{equation}
\rho_{-} = -\frac{1}{4\pi a^4}.
\end{equation} 
Thus there is some strong ground to believe that the negative energy appearing in the complex extension of Milne model is of the nature of Casimir energy. However, one must admit that a much deeper justification for this, on the basis of a true quantum field or similar theory, is needed here.

This caveat is there also when the energy of the field $\phi$ is assumed to contribute to the energy of matter/radiation and the time-variable dark energy. Similar is the case for the creation of matter/radiation from dark energy, as envisaged in this model. One can only view these  predictions as providing outlines of a future broader theory. But since we have the result from quantum cosmology that there is exact classical to quantum correspondence for this new model,  the above  pointers from the classical theory  can be expected to be in the right direction.

\section*{Acknowledgements}
I thank Professor K. Babu Joseph  for valuable advices,  Ninan Sajeeth Philip for some useful references and Kiran Mathew for technical help in manuscript preparation.

\label{lastpage}
\end{document}